\documentclass{PoS}

\def\s{s\,}
\def\bs{\bar s}

\def\noeq#1{(\ref{#1})}
\def\1eq#1{Eq.~(\ref{#1})}

\def\2eqs#1#2{Eqs.~(\ref{#1}) and~(\ref{#2})}
\def\3eqs#1#2#3{Eqs.~(\ref{#1}),~(\ref{#2}) and~(\ref{#3})}

\title{Gauge theories with non-trivial backgrounds}

\ShortTitle{Gauge theories with non-trivial backgrounds}

\author{\speaker{Daniele Binosi}\\
        European Centre for Theoretical Studies in Nuclear Physics
        and Related Areas (ECT*), \\and Fondazione Bruno Kessler,\\
        Villa Tambosi, Strada delle Tabarelle 286, I-38123 Villazzano (TN)  Italy.\\
        E-mail: \email{binosi@ectstar.eu}}

\abstract{We review our most recent results in formulating gauge theories in the presence of a background field on the basis of symmetry arguments only. In particular we show how one can gain full control over the dependence on the background field of the effective action, and how the so-called background field method emerges naturally from the requirement of invariance under the BRST and antiBRST symmetries.}

\FullConference{QCD-TNT-III-From quarks and gluons to hadronic matter: A bridge too far?,\\
		2-6 September, 2013\\
		European Centre for Theoretical Studies in Nuclear Physics and Related Areas (ECT*), Villazzano, Trento (Italy)}

\begin{document}

\section{Motivation}

When working in a gauge theory, it is quite common to face the problem of having to consider the quantum fluctuations of the gauge field around some non-trivial background, the properties of which are generally determined by the physical problem one is considering.

The standard approach is then to split the gauge field $A_\mu$ as
\begin{equation}
A^a_\mu=\widehat{A}_\mu^a+Q_\mu^a,
\label{classicalbq}
\end{equation}
where  $Q$ represents the quantum fluctuations whereas $\widehat{A}$ is the background field; then one carries out the path integral over the $Q$ variables. This is, {\it e.g.}, what 't~Hooft did in his classic computation of the one-loop effective action evaluated around an instanton configuration~\cite{Hooft:1976fv}.  A second relevant example, in which the methods discussed here are especially useful, is provided by the Color Glass Condensate (CGC)~\cite{Iancu:2003xm} in which the different scales present in the problem allows one to describe the physics by integrating out  ``semi-fast'' modes (representing the $Q$ part above) keeping fixed a background that is split into a soft part
plus a ``fast'' classical component. The latter is in turn determined by static color sources, associated to some statistical distribution whose modification under the integration of the semi-fast modes is described by the CGC evolution equations (for details see~\cite{Quadri:QCDTNT-III}).

Another interesting physical situation is the one encountered in the so-called Background Field Method (BFM)~\cite{Abbott:1980hw}, in which case one considers the background $\widehat{A}$ as an unspecified source to be set to zero after taking the appropriate number of derivatives of the vertex functional ({\it e.g.}, two if we are interested in the correlator of two background gauge bosons). In such case, a residual gauge invariance with respect to the gauge transformations of the background field is left, which can be exploited to simplify all kind of calculations both in the perturbative as well as in the non-perturbative regime, or even used as a powerful prescription to construct gauge invariant and renormalization groups invariant quantities~\cite{Binosi:2009qm}.

Until very recently, however, the role played by symmetry in the formulation of theories in the presence of a background field has been largely unexplored. What I will summarize here is the work that we have been developing in a recent series of papers~\cite{Binosi:2011ar,Binosi:2012pd,Binosi:2012st,Binosi:2013cea} where we have described some particularly powerful tools and methods appropriate to introduce in a gauge theory a background field in such a way that the complete dependence of the theory on the latter is determined through symmetry arguments alone. I will pay particular attention to the implication these methods have at both the theoretical and phenomenological level. 

\section{The $\Omega$ source}

Let's first start from conventional perturbation theory, where indeed the control on the background quantum splitting can be achieved through a symmetry requirement alone. Specifically, one extends the ordinary BRST symmetry to encompass the shift symmetry of the background field, by combining the latter in a BRST doublet together with a source $\Omega$~\cite{Grassi:1995wr,Becchi:1999ir,Ferrari:2000yp}:
\begin{equation}
\s\widehat{A}^a_\mu=\Omega^a_\mu;\qquad \s\Omega^a_\mu=0.
\end{equation}

Even though the geometrical meaning of this source is somewhat obscure at this stage, this turns out to be a very good idea, as it provides an extended Slavnov-Taylor (ST) identity that enforces a set of relations --the so-called background-quantum (BQ) identities~\cite{Grassi:1999tp,Binosi:2002ez}-- that must be fulfilled by the background amplitudes; in addition the residual gauge invariance can be also expressed in functional form. Thus one has the following identities 
\begin{eqnarray}
&&\int\!\mathrm{d}^4x\left[\Gamma_{A^{*a}_\mu}\Gamma_{A^\mu_a}+\Gamma_{c^*_a}\Gamma_{c^a}+b^a\Gamma_{\bar c^a}+\Omega^a_\mu\Gamma_{\widehat{A}^\mu_a}
\right]=0;\qquad \Gamma_\varphi=\frac{\delta\Gamma}{\delta \varphi},\nonumber \\
&&-\partial_\mu\frac{\delta\Gamma}{\delta \widehat{A}^a_\mu}+f^{abc}\widehat{A}^c_\mu\frac{\delta\Gamma}{\delta \widehat{A}^b_\mu}+\sum_\varphi f^{abc}\varphi^c\frac{\delta\Gamma}{\delta\varphi^c}=0;\qquad \varphi=\{Q,c,\bar c,b,\Omega,A^*,c^*\},
\label{STI-WI}
\end{eqnarray} 
where a `$*$' indicates the antifield of the corresponding field, and $b$ is the Nakanishi-Lautrup multiplier.
Given this setting, we proved recently a theorem stating that~\cite{Binosi:2011ar}:
\vspace{.5cm}

\noindent{\it The extended ST identity uniquely fixes the dependence of the functional background field.} 
\vspace{.25cm}

The general proof relies on cohomological techniques, but the way it proceeds can be described in a simplified way as follows.
Take a derivative of the ST identity with respect to the source $\Omega$, which is set to zero afterwards; one then obtains a first order functional differential equation for the effective action
\begin{equation}
\Gamma_{\widehat{A}^\mu_a}(x)=\int\!\mathrm{d}^4y\,\Gamma_{\Omega^a_\mu A^{*b}_\nu}(x,y)\Gamma_{A^\nu_b}(y);\qquad \widehat{A}^a_\mu\ne 0.
\end{equation}
Then assume that one can find a functional ${\cal G}$ such that 
\begin{equation}
\frac{\delta{\cal G}^b_\nu(x)}{\delta\widehat{A}^a_\mu(y)}=\Gamma_{\Omega^a_\mu A^{*b}_\nu}(x,y).
\end{equation}
Then $\Gamma[A,\widehat{A}]=\Gamma[0,A-{\cal G}]$ is the complete solution of the above equation\footnote{The difficult part of the proof is to show that there are no obstructions to the integration of the functional equation defining ${\cal G}$; this is far from obvious, as one soon realizes that a naively expected exponentiation of the solution fails~(see~\cite{Binosi:2011ar} for details).}. Therefore the full dependence on the background field is generated through the redefinition of the quantum gauge field 
\begin{equation}
A^a_\mu=\widehat{A}_\mu^a+Q_\mu^a\quad\longrightarrow\quad A^a_\mu=
\underbrace{\widehat{A}_\mu^a-{\cal G}^a_\mu}_{V^a_\mu} +Q_\mu^a,
\end{equation}
which generalizes the classical quantum/background splitting~\noeq{classicalbq}. 

It is at this point when things starts getting interesting. Specifically, assume that we are able to prove that in a given theory (possibly within suitable approximations) the function $\Gamma_{\Omega A^*}$ vanishes: $\Gamma_{\Omega A^*}=0$. Then we know that the background/quantum splitting stays classical and the background field does not get deformed by quantum corrections. Therefore if $\widehat{A}$ is a solution of the classical equations of motion it will be so even after quantum corrections are fully taken into account. This is the case in the CGC effective theory~\cite{Quadri:QCDTNT-III}. 

However the CGC case represents an exception rather than the rule: in general to lowest order in the quantum corrections, a background field will be deformed according to
\begin{equation}
{V^a_\mu}(x)=\widehat{A}_\mu^a(x)+\int\!\mathrm{d}^4y\,\left.\Gamma^{(1)}_{\Omega^a_\mu A^{*b}_\nu}(x,y)\right\vert_{\widehat{A}=0}\widehat{A}^b_\nu(y),
\end{equation}
where, as explicitly indicated, the deformation function $\Gamma_{\Omega A^*}$ is now evaluated at zero background field.

This applies, {\it e.g.}, in the case of an instanton background, where the (singular gauge) instanton profile gets modified according to~\cite{Binosi:2012pd}
\begin{eqnarray}
{V^a_\mu}(x)&=&\bar\eta_{\mu\nu}^a x_\nu f(\lambda)\nonumber \\
f(\lambda)&=&\frac1{\rho^2}\left\{\left[2-3\frac{g^2}{8\pi^2}\left(1+\log\rho\mu\right)\right]\frac1{\lambda^2(1+\lambda^2)}\right.\nonumber \\
&+&\left.3\frac{g^2}{8\pi^2}\frac1{\rho^2}\left[-\frac{\gamma_{\mathrm E}-\log2}{\lambda^2(1+\lambda^2)}-\frac{\log\lambda}{\lambda^2}+\frac{1+\lambda^4}{2\lambda^4(1+\lambda^2)}\log(1+\lambda^2)\right]\right\},
\end{eqnarray}
with $\bar\eta$ the t' Hooft symbol, $\rho$ the instanton size, $\lambda=r/\rho$ and, finally,  $\gamma_{\mathrm E}$ the Euler-Mascheroni constant, $\gamma_{\mathrm E}=0.57721\cdots$.

We then see that when quantum corrections are taken into account the self-dual and pure gauge character of the instanton solution are lost; on the other hand one gets supporting evidence that such corrections lead to a log enhancement for small and large size instantons and therefore to a suppression of the instanton density in these regimes, which is incidentally what the simulations on the lattice indicate too\footnote{Notice that this has to be considered at most indicative, as it is exactly in these regimes that perturbation theory breaks down.}. However, the infrared disease of instanton calculus is not cured, as the density diverges when $r\to0$: one still needs a dynamically generated gluon mass~\cite{Cornwall:1981zr} to avoid this.

\section{Beyond the $\Omega$}

The BFM formulation discussed so far still needs the presence of dynamical ghost fields. We would then like to know if the presence of these fields is unavoidable, or rather there exists a  formulation in which they are not necessary at all.

The first step in this direction is to rewrite the ST identity in terms of the so-called Batalin-Vilkoviski (BV) bracket of the theory, which is defined as\footnote{Here and in what follows all the derivatives are  left derivatives; for the graded properties of the bracket see~\cite{Gomis:1994he}.}~\cite{Gomis:1994he}
\begin{equation}
\{X,Y\} = \int\!{\mathrm d}^4x \sum_\varphi
\left[ (-1)^{\epsilon_{\varphi} (\epsilon_X+1)}
X_\varphi  Y_{\varphi^*}- (-1)^{\epsilon_{\varphi^*} (\epsilon_X+1)}
X_{\varphi^*} Y_{\varphi}
\right],
\label{BVbracket}
\end{equation} 
where $\epsilon_\varphi$, $\epsilon_{\varphi^*}$ and $\epsilon_X$ represent the statistics of the field $\varphi$, the corresponding antifield $\varphi^*$ and the functional $X$ respectively. 

Then one has that the ST identity in Eq.~(\ref{STI-WI}) can be rewritten as
\begin{equation}
\int\!{\mathrm d}^4x\, \Omega^a_\mu(x)
\Gamma_{\widehat A^a_\mu}(x) = 
- \frac{1}{2}\, \{\Gamma,\Gamma\}.
\label{m.1}
\end{equation}
Next, take a derivative with respect to $\Omega$ of the equation above, and set this source to zero, to obtain~\cite{Binosi:2012pd}
\begin{equation}
\left.\Gamma_{\widehat A^a_\mu}(x)\right\vert_{\Omega=0}=-\{\underbrace{\Gamma_{\Omega^a_\mu}(x)}_{\Psi^a_\mu(x)},\Gamma\}\Big\vert_{\Omega=0}.
\end{equation}

We see here a rather peculiar equation, which states that the derivative of the vertex functional with respect to the background field equals the  effect of an infinitesimal canonical  transformation (with respect to the BV bracket) on the vertex functional itself\footnote{That the variables $\varphi$ and $\varphi^*$ are canonical with respect to the BV bracket, can be readily seen by computing $\{\phi_i(x),\phi_j(y)\}=\{\phi^*_i(x),\phi^*_j(y)\}=0$ and
$\{\phi_i(x),\phi^*_j(y)\}=\delta_{ij}\delta^4(y-x)$.}. 
This fact allows us to reformulate the problem of finding the complete dependence of the effective action in the following terms~\cite{Binosi:2012st}:
\vspace{.5cm}

\noindent{\it Find a canonical mapping between the old variables $\{\varphi,\varphi^*;\widehat{A}\}$ and some new (canonical) variables $\{\Phi,\Phi^*\}$ such that the ST identity~(\ref{m.1}) written in these new variables is automatically satisfied, or\footnote{Incidentally, these equations explain the failure of the naive exponentiation noticed before, for the generating function $\Psi$ depends explicitly on the background field}
\begin{eqnarray}
\frac{\delta\Phi(y)}{\delta \widehat{A}^a_\mu(x)}&=&\frac{\delta\Psi^a_\mu(x)}{\delta \Phi^*(y)}=\{\Phi(y),\Psi^a_\mu(x)\},\nonumber \\
\frac{\delta\Phi^*(y)}{\delta \widehat{A}^a_\mu(x)}&=&-\frac{\delta\Psi^a_\mu(x)}{\delta \Phi(y)}=\{\Phi^*(y),\Psi^a_\mu(x)\}.
\label{Phieqs}
\end{eqnarray}}
\vspace{.25cm}

Then, since the BV bracket does not depend
on either $\widehat A$ or $\Omega$, such canonical mapping would control the full dependence of the effective action $\Gamma$
on the background field; and this would happen not only at the level of the counterterms  of $\Gamma$, but rather for the full 1-PI Green's functions, thus giving control even over the non-local dependence of $\Gamma$ on the background.  

The way to actually solve the equations~(\ref{Phieqs}) is through a a field theoretical generalization of the classical concept of a Lie transform introduced long ago by Deprit in classical mechanics~\cite{Deprit:1969aa}. To this end, one defines the operator~\cite{Binosi:2012st}  
\begin{equation}
\Delta_{\Psi^{a}_\mu(x)}=\{\cdot,\Psi^{a}_\mu(x)\}+\frac\delta{\delta \widehat{A}^a_\mu(x)},
\end{equation}
in which the first term constitutes a graded generalization of the classical Lie derivative, while the second term accounts for the explicit background dependence of the generating function. Then the sought for canonical transformation is given in terms of a formal power series in the background field~\cite{Binosi:2012st}:
\begin{eqnarray}
\Phi(x) &=& \sum_{n \geq 0}
\frac{1}{n!}  \int_1\! \cdots\! \int_n\!
\widehat A_1\cdots\widehat A_n  
\left [ \Delta_{\Psi_n}\! \cdots \Delta_{\Psi_1} \phi(x) \right ]_{\widehat A=0},\nonumber \\
\Phi^*(x) &=& \sum_{n \geq 0}
\frac{1}{n!}  \int_1\! \cdots\! \int_n\!
\widehat A_1\cdots\widehat A_n  
\left [ \Delta_{\Psi_n}\! \cdots \Delta_{\Psi_1} \phi^*(x) \right ]_{\widehat A=0}.
\label{cantras}
\end{eqnarray}
Therefore, if we know the Green's functions of a given theory for the quantum fields {\it only} (that is, satisfying the ST identity at zero background field), the full dependence of the background field can be obtained through the canonical transformation~(\ref{cantras}). 

The important point to notice however is that  the canonical transformation~\noeq{cantras} can also be written in a (gauge-invariant) model where the ghosts are replaced by external classical anticommuting sources: thus our original question is answered in the positive, as this formulation of the BFM in terms of canonical transformations shows that {\em dynamical} ghosts need not be present. 
This, in turn, overcomes the  absence at the non-perturbative level of the BRST symmetry, thus avoiding the (in)famous Neuberger 0/0 problem, whence paving at the same time the implementation of the BFM on the lattice.  
Indeed, it is not difficult to prove that the minimization of the functional~\cite{Binosi:2012st,Cucchieri:2012ii}
\begin{equation}
F[g]=-\int\!{\mathrm d}^4x\,{\mathrm Tr}\,(A^g_\mu-\widehat{A}_\mu)^2;\qquad
A_\mu^g = g^\dagger  A_\mu g - i \partial_\mu g^\dagger g,
\end{equation} 
over the gauge group elements $g$ yields the background Landau gauge condition\footnote{We define the covarinat derivative according to ${\cal D}_\mu=\partial_\mu-igA_\mu$, or in components ${\cal D}^{ab}_\mu=\partial_\mu\delta^{ab}+f^{acb}A^c_\mu$; correspondingly the background covariant derivative is obtained from the previous formulas through the replacement $A\to\widehat{A}$.} $\widehat{\cal D}^\mu(A^g_\mu-\widehat{A}_\mu)=0$. Then {\it on the minimum} the mapping $A\to A^g(A,\widehat{A})-\widehat{A}$ defines the action of the canonical transformation on the gauge field, thus generalizing non perturbatively the background quantum splitting.

\section{BRST+antiBRST=BFM}

The last result I would like to present is a recent finding~\cite{Binosi:2013cea} that incidentally explains the origin of the $\Omega$ source we started from.
As this has to do with the concept of the antiBRST symmetry, which might be something not everybody is acquainted with, let me start briefly reviewing it.

In Yang-Mills theories the ghost and the antighost fields play a very asymmetric role. Specifically, whereas $c$ replaces the gauge fixing parameter and its behavior is fixed by the cohomology of the Lie algebra, $\bar c$ and $b$ are just Lagrange multipliers for the gauge fixing condition and its BRST transform respectively. In addition, $\bar c$ is not the Hermitian conjugate of $c$ and, as such, it satisfies a different equation of motion.

Nevertheless it was found long ago~\cite{Curci:1976bt} that it is still possible to introduce a nilpotent symmetry in which the ghost roles are exchanged. Its defining transformations are obtained from the BRST ones by the replacements $c\leftrightarrow\bar c$ and $b\leftrightarrow\bar b$:
\begin{equation}
\bs A^a_\mu={\cal D}^{ab}_\mu \bar c^b;\qquad
\bs \bar c^a=-\frac12f^{abc}\bar c^b\bar c^c;\qquad
\bs c^a=\bar b^a;\qquad
\bs \bar b^a=0.
\label{antiBRST}
\end{equation}
Requiring the algebra to close enforce the additional transformations
\begin{equation}
s\bar b^a=f^{abc}\bar b^b c^c;\qquad
\bs b^a=f^{abc} b^b\bar c^c,
\label{extra}
\end{equation}
while demanding the nilpotency of the product of the BRST and antiBRST operators, that is $\{s,\bar s\}=0$, implies a constraint relating $b$ and $\bar b$:
\begin{equation}
\bar b^a=-b^a-f^{abc}c^b\bar c^c,
\end{equation}
which, while consistent with~\1eq{extra}, explicitly breaks an extra Sp[2] symmetry $(c,b)\leftrightarrow(\bar c,\bar b)$ which would be otherwise present.

Suppose now that we want to render the usual Yang-Mills action BRST and anti-BRST invariant. In that case one has to consider also the non-trivial $s\bar s$ transformations which read
\begin{equation}
s\bs A^a_\mu={\cal D}^{ab}_\mu b^b+f^{abc}\left({\cal D}_\mu^{bd}c^d\right)\bar c^c;\qquad
s\bs c^a=s\bar b^a;\qquad
s\bs \bar c^a=-\bs b^a.
\end{equation}

Thus, overall, a BRST-antiBRST invariant Yang-Mills theory requires the introduction of eight sources~\cite{Binosi:2013cea}: the usual antifields $\{A^*,c^*\}$, the antifields of the antiBRST symmetry $\{A^\#,b^\#,c^\#,\bar c^\#\}$, and, finally, two BRST-antiBRST sources, for which we will choose the suggestive notation $\{\widehat{A},\widehat{c}\}$. Then, before gauge fixing, the Yang-Mills invariant action reads
\begin{equation}
S_{\mathrm I}=S_{\mathrm{YM}}+\sum_\varphi\!\int\mathrm{d}^4x\,\left(\varphi^*s\,\varphi+\varphi^\#\bar s\,\varphi+\widehat{\varphi}s\bar s\,\varphi\right),
\end{equation} 
where 
\begin{equation}
s\varphi^*=\bs\varphi^*=0;\qquad
s\varphi^\#=\bs\varphi^\#=0;\qquad 
s\widehat{\varphi}=\varphi^\#;\qquad
\bs\widehat{\varphi}=-\varphi^*,
\end{equation}
with the exception of $sb^\#=\bar c^\#$ and $\bs b^\#=0$.

Now let's add to the above action the usual $R_\xi$ gauge fixing term ${\cal F}^a_\mu=\partial^\mu A^a_\mu$; observing that
\begin{equation}
s\left[\bar c^a\widehat{\cal F}^a-\frac\xi2\bar c^a b^a\right]
=s\left[\bar c^a{\cal F}^a-\frac\xi2\bar c^a b^a\right]+\widehat{A}^\mu_a (s\bar s A^a_\mu)+\Omega^\mu_a(\bar s A^a_\mu),
\end{equation}
one finds that once we identify $\Omega$ with $A^\#$ we have discovered the following~\cite{Binosi:2013cea}:
\vspace{.5cm}

\noindent{\it Requiring the invariance of an $R_\xi$ gauge-fixed Yang-Mills action under both the BRST and antiBRST symmetry, is equivalent to quantizing the theory within the BFM $R_\xi$ gauges\footnote{Similarly, when scalars are present, BRST-antiBRST invariance implies quantization in the 't Hooft background gauge.}.}

\vspace{.5cm}

\noindent Thus, on the one hand the source $\Omega$ appears naturally as the source of the antiBRST variation of the gauge field $A$ in an BRST-antiBRST invariant context; on the other hand, the background field is the source
of the BRST-antiBRST variation $s \bar s A$ of the gauge field\footnote{Notice however that $\widehat{c}$ cannot be interpreted as a background 
for the ghost $c$, since it has ghost number $-1$; 
it is also clear that it is not a background for the antighost field, as a shift of the latter field would lead to totally different couplings with respect to the ones that are generated for the source $\widehat{c}$.}.

We conclude by observing that as BRST invariance implies the existence of a local ghost equation, antiBRST invariance furnishes a local antighost equation, which, for any value of the gauge-fixing parameter reads
\begin{equation}
\Gamma_{c^a}+f^{abc}\Gamma_{b^b}\bar c^c+\xi\Gamma_{b_a^\#}-\widehat{\cal D}^{ab}_\mu\Gamma_{A^{\#b}_\mu}-f^{abc}\widehat{c}^b\Gamma_{c^\#_c}-f^{abc}b^{\#}_b\Gamma_{\bar c^\#_c}={\cal D}^{ab}_\mu A^{*\mu}_b+f^{abc}c^*_b c^c.
\end{equation}
Once combined with the local ghost equation one can fully constrain the ghost two point sector of the theory, and in particular obtain a generalization of the usual formula relating the ghost dressing function in the Landau gauge with certain auxiliary Green's functions~\cite{Kugo:1995km,Grassi:2004yq,Aguilar:2009pp} appearing naturally in the context of the so-called PT-BFM framework~\cite{Binosi:2007pi,Binosi:2008qk}.

\section{Epilogue}    

We have reached a quite complete understanding of the BFM both when it is employed for physical backgrounds 
as well as when the background is treated as an auxiliary
external source, allowing to establish a background Ward
identity that can be used to greatly simplify perturbative and nonperturbative computations.
In both cases we have shown that many interesting results can be derived on the basis of symmetry arguments alone. Perhaps the most surprising result of the studies presented here is the fact that the BFM turns out to be ``emergent'' from the requirement of BRST-antiBRST invariance: as such, Yang-Mills theories quantized in the BFM $R_\xi$  type of gauges should be regarded as the most symmetric realization of these theories as they encode invariance under both BRST and antiBRST symmetry.

%\bibliography{../../Papers/PinchTechnique/Bibliography/bibliography}
%\bibliographystyle{JHEP}

\providecommand{\href}[2]{#2}\begingroup\raggedright\endgroup

\end{document}